\documentclass[twocolumn,showpacs,prl]{revtex4}
\usepackage{amsmath}
\usepackage{mathrsfs}
\usepackage{graphicx}

\begin{document}
\draft
\title{Conditional relaxation of a charge state under continuous weak measurement }%

\author{Gyong Luck Khym$^{1,2}$}
\author{Kicheon Kang$^1$ }
\email{kicheon.kang@gmail.com}

\affiliation{$^1$Department of Physics, Chonnam National
University, Gwangju 500-757,  Korea \\
$^2$Department of Physics, Pohang University of Science and Technology, Pohang 790-784, Korea}%

\date{\today}

\begin{abstract}
 We investigate the conditional evolution of a charge state coupled to
 a mesoscopic detector under continuous weak measurement.
 The state suffers relaxation into
  a particular state with a definite charge when electrons in a particular
output lead are monitored in the detector.  The process of the conditional relaxation is not restricted
 by the shot noise of the detector, unlike the case of the back-action dephasing.
 As a result, the relaxation of conditional evolution is much faster
than the current-sensitive part of dephasing. Furthermore,
 the direction of the relaxation depends
 on the choice of the output lead.
 We propose that these properties can be verified in a two-path
interferometer containing a quantum dot capacitively coupled to a detector. In this setup,
 the current-current correlation between the interferometer and the
 detector reveals characteristic features of conditional relaxation.

\end{abstract}


\maketitle

\newcommand \tr {{\rm Tr}}

The quantum measurement problem continues to attract interest because
a measurement process inevitably causes the ``wave function reduction"
that cannot be described in terms of the Schr\"odinger equation~\cite{greenstein06}.
Mesoscopic physics has recently progressed into a stage that enables us
to treat this issue. In particular,
a quantum dot entangled with a mesoscopic conductor undergoes ``back-action
dephasing" experimentally realized~\cite{buks98,
sprinzak00,chang08}. This dephasing has also been a subject of
intensive theoretical
investigation~\cite{aleiner97, gurvitz97, levinson97, hacken98,
stodolsky, butt-martin, korotkov-averin, silva01, pilgram,
averin05, kang05, khym06j,khym06ja}.
The back-action dephasing
can be understood in terms of the possibility of acquiring charge-state
information. However, it is important to note that the actual
measurement has not been performed for the dephasing process. It only
refers to the possibility of measurement and is a result of averaging
over all possible measurement outcomes.
On the other hand, a quantum
measurement performed on the detector brings about a sudden reduction of
the charge state (or the ``wave function collapse")~\cite{nielson00}. Continuous measurement on a particular outcome of the detector state results in
an evolution of the charge state in a way that depends on the choice of measurement outcome.

The system under study is schematically drawn in Fig.~1.
 A quantum point contact (QPC) adjacent to the target system (usually a quantum dot) can be used as a charge detector through
the charge-sensitivity of the detector
current~\cite{field93,buks98,chang08}. The information of the charge state is transferred
 to the detector in the form of a quantum entanglement. There are two possible outcomes of measurement in
 the QPC detector, that is, transmission and reflection, for each of the
detector electrons. Transport through a quantum dot
 coupled to a QPC detector depends on what detector output current is observed~\cite{sukhorukov07}, demonstrating the conditional statistics.
 The nature of electron transport in the detector is stochastic because of
random partitioning at the QPC. The stochastic evolution of
 the charge state under this random selection of the detector state has been studied before~\cite{korotkov99,averin06}.

 In our study, in contrast, we investigate the evolution of the charge state of the
target system with the condition that only one particular lead of the
detector is intentionally monitored.  Our main observations are:
(1) The initial state given as a coherent superposition of
 two different charge states is relaxed to the one of the fixed charge state. The direction of the relaxation depends on the
 choice of measurement on the detector. That is, the charge state is relaxed
 to $|0\rangle$ (state without an extra charge) conditioned on the selection
 of the detector electron at $T$.
On the other hand, the charge state is relaxed
 to $|1\rangle$ (state with an extra charge) when electrons are continuously
selected at lead $R$. (2) The relaxation rate is the same in both cases
and is
proportional to the charge sensitivity of
 the detector transmission. The relaxation rate is much larger than the
current-sensitive part of the dephasing rate, which can be regarded as
a manifestation of nonlocality in a measurement process.

We propose an experimental setup which can be used to verify this
conditional relaxation. In order to monitor the state of the target system,
we introduce a quantum dot embedded in a two path interferometer. The
electronic Mach-Zehnder interferometer
 with a quantum Hall edge channel~\cite{ji03} is an ideal system for this
purpose, but the conventional type of
Aharonov-Bohm interferometer~\cite{schuster97} can also be used. For charge detection,
a QPC is considered which is capacitively coupled to the quantum dot. We show that, while the current oscillation amplitude
in the interferometer is directly related
to dephasing via entanglement, the cross correlation of the currents (between a lead of the interferometer and the other
in the detector) reveals the characteristic features of the conditional relaxation.

\begin{figure}[b]
\includegraphics[width=75mm]{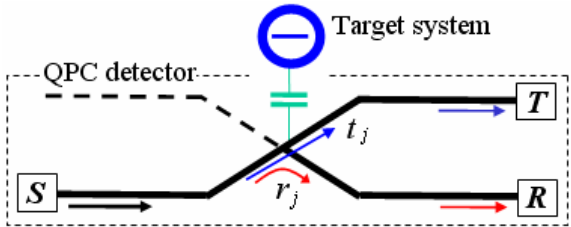}
 \caption{\label{fig1}
 A schematic diagram of a target system coupled to a quantum point contact detector.
 The state information is encoded in the charge-dependent reflection and transmission amplitudes,
 $r_j$ and $t_j$, respectively, in the detector ($j=0,1$).
 }
\end{figure}

Initially, the charge state of the target system is in general given as
a linear
superposition, $a_0 |0\rangle + a_1 |1\rangle$, of two different
charge states, $|0\rangle$ and $|1\rangle$, respectively. Electron
scattering via QPC detector is affected by the state of the target
system and is accounted for in the scattering matrix (for $j=0,1$)
 \begin{eqnarray}
 S_{pc} =   (\delta_{j0} +
 \delta_{j1} ) \left(
        \begin{array}{cc}
           r_{j} & t'_{j} \\
       t_{j} & r'_{j}
     \end{array}
       \right),  \label{qpc-smtx2}
\end{eqnarray}
where its elements depend on the charge state $|j\rangle$.
For a detector bias $V_{det}$,
the average number of electrons injected into the QPC during the time
interval $t$ is $n=eV_{det}t/h$.  We are
interested in the limit of continuous measurement, that is,
$n\gg1$, and neglect the energy dependence of the matrix elements.

The electron creation(annihilation) of energy $\epsilon$ at lead $l$ ($l=S,T,R$) is
represented by the operator $c^{\dagger}_{l}(\epsilon)$
($c_{l}(\epsilon)$). The initial state is a direct
product of the charge state $ a_0|0\rangle + a_1 |1\rangle$ and
the detector state
 $\prod_{0<\epsilon\leq eV_{det}} c^{\dagger}_S(\epsilon)|F
 \rangle$ where $|F\rangle$ is the Fermi sea of electrons
with $\epsilon<eV_{det}$.
 Upon interaction of $n$ detector electrons with the charge state, the two subsystems get entangled as
\begin{subequations}
\begin{equation}
  |\Psi \rangle =
  a_0 |0\rangle \otimes \Big[ \prod_{\epsilon}\chi_0^\dagger(\epsilon)\Big]|F\rangle
  +a_1 |1\rangle \otimes\Big[\prod_{\epsilon} \chi_1^\dagger(\epsilon)\Big]|F\rangle,
 \label{entangle}
 \end{equation}
where the energy interval $0\leq \epsilon \leq eV_{det}$ is counted.
 $\chi_j^\dagger(\epsilon)$ ($j=0,1$) creates a charge-state-dependent detector electron:
\begin{eqnarray}
\chi_j^\dagger(\epsilon) =  r_j c^{\dagger}_R (\epsilon)
 + t_j c^{\dagger}_T (\epsilon). \label{state-of-det}
\end{eqnarray}
\end{subequations}

Dephasing of the charge state induced by this type of entanglement is now well
understood~\cite{aleiner97, gurvitz97, levinson97, hacken98,stodolsky,butt-martin, korotkov-averin, silva01, pilgram,
averin05, kang05}. First, we briefly review the dephasing properties of the charge state.
 The charge state is
 described by a reduced density matrix  $\rho= \tr_{det}\big[ |\Psi \rangle
\langle \Psi| \big]$, where $\tr_{det}[\cdots]$ sums over the detector's
degrees of freedom. From this, we can find the time evolution of the
density matrix elements,
\begin{eqnarray}
 \ln \rho_{jj^{\prime}}(t) = \ln \rho_{jj^{\prime}}(0) +  \sum_{0<\epsilon\leq eV_{det}} \ln{[
 \nu_{jj^{\prime}}(\epsilon)], }\label{Dephasing}
 \end{eqnarray}
where
 $\nu_{jj^{\prime}}(\epsilon) = r_{j^{\prime}}^{*} r_{j} +t_{j^{\prime}}^{*} t_{j}$ is the quantity that accounts for the effect of charge detection. The initial density matrix is  $\rho_{jj'}(0)=a_ja_{j'}^*$.
Eq.~(\ref{Dephasing}) indicates that the diagonal components are
unchanged, but the off-diagonal terms decay as a function of time leading to dephasing.
 In the limit of $t \gg h/eV_{det}$ with $|\nu_{01}(\epsilon)|\sim 1$ (weak continuous measurement), we obtain the asymptotic relation
 $|\rho_{01}(t)| = |\rho_{01}(0)| \exp(-\Gamma_{dep}t)$
 where the dephasing rate $\Gamma_{dep}$ is given by
 $\Gamma_{dep} = -\int h^{-1}{d \epsilon} \ln |\nu_{01}(\epsilon)|$.
 Due to the condition of weak measurement ($|\nu_{01}(\epsilon)|\sim 1$),
 $\Gamma_{dep}$ can be expanded in terms of the change in the
 transmission probability $\Delta {\cal T} = {\cal T}_0-{\cal T}_1$
(${\cal T}_j\equiv |t_j|^2$) and the change in the relative scattering phase
 $\Delta \phi \equiv \arg(t_0/r_0
 )-\arg(t_1/r_1)$. We find
\begin{eqnarray}
 \Gamma_{dep}
  = \frac{eV_{det}}{8 h }   \frac{(\Delta {\cal T})^2}{{\cal T}(1-{\cal T})} +   \frac{eV_{det}}{2 h} {\cal T}(1-{\cal T})(\Delta
 \phi)^2 ,  \label{Dephasing-rate}
\end{eqnarray}
 where $ {\cal T} = ({\cal T}_0+{\cal T}_1)/2$.

Next, we discuss our main observation of the conditional evolution of
the charge state. In the above, we have described dephasing
of the charge state by its entanglement with the detector electrons.
Actual measurement for the detector is not performed for dephasing of the charge state.
In contrast, we can monitor the charge state of the target system
under continuous selection of detector electrons at a particular lead. (This corresponds to a continuous projective measurement.)
The conditional state is obtained by projecting the total state
into a state with a specific
outcome of measurement and  renormalizing the reduced wave
function~\cite{nielson00}. It is important to note that, under this circumstance, the charge state is not entangled with the detector state, and remains as a pure state, as long as the initial state of the target system is pure. In the particular setup of Fig.~1, there are two
possible outcomes for measurement on the detector, that is,
transmission and reflection, for each of the detector electrons. So, there
are two different ways of continuous projection for the detector outputs. This measurement is given by the operator
\begin{subequations}
\label{eq:My}
\begin{equation}
 M_y =  \large[  \langle \Psi  |  y \rangle \langle  y| \Psi
    \rangle \large]^{-1/2}  |y\rangle\langle y|,
\end{equation}
 where  $| y\rangle =\Big[\prod_{\epsilon} c^{\dagger}_y (\epsilon)
\Big] |F\rangle$. The case $y=R$ ($y=T$) corresponds to a continuous
projection of the detector state onto lead $R$ ($T$).
The corresponding state of the composite system evolves as
\begin{equation}
  |\Psi\rangle \rightarrow M_y |\Psi\rangle
  = |\psi^y(t)\rangle \otimes |y\rangle.
\end{equation}
Clearly, the two subsystems are disentangled upon the measurement
as a result of
the ``wave-function collapse". The conditional state of the target system is
\begin{equation}
   |\psi^y(t)\rangle =  A_y(t)|0\rangle
 + B_y(t)| 1 \rangle,
\end{equation}
where the coefficients $A_y(t)$ and $B_y(t)$ satisfy the relations
\begin{eqnarray}
   {B_R(t) \over  A_R(t)}  =  {a_1 \over a_0} \prod_{\epsilon} {r_1 \over r_0   },~
   {B_T(t) \over  A_T(t)}  =  {a_1 \over a_0} \prod_{\epsilon} {t_1 \over t_0}.
\end{eqnarray}
In the asymptotic limit ($t\gg h/(eV_{det}{\cal R}_0)$ for $y=R$,
$t\gg h/(eV_{det}{\cal T}_0)$ for $y=T$), we find
\begin{equation}
 {{B_R(t)} \over {A_R (t)}} = e^{(\Gamma^R_{rel}/2+i\xi_R)t}  {{a_1}\over {a_0} }, \;\;
 {{B_T(t)} \over {A_T (t)}} = e^{(-\Gamma^T_{rel}/2+i\xi_T)t}  {{a_1}\over {a_0} },
\end{equation}
where the relaxation rates are
\begin{eqnarray}
 \Gamma^R_{rel} &=& 2 {\cal R}_0 \int h^{-1} d\epsilon\ln |r_1/r_0| , \\
 \Gamma^T_{rel} &=& -2 {\cal T}_0 \int h^{-1} d\epsilon\ln |t_1/t_0| .
\end{eqnarray}
Here ${\cal R}_j = |r_j|^2$ is the reflection probability.
The measurement also induces the phase shifts $\xi_R ={\cal R}_0 eV_{det}\arg(r_1/r_0)/h$ and
  $\xi_T = {\cal T}_0 eV_{det}\arg(t_1/t_0)/h$, respectively.
Imposing conditions for weak measurement, ($\Delta {\cal T} / 
{\cal R}_0
\ll 1 $ for $y=R$ and $\Delta {\cal T} / {\cal T}_0 \ll 1$ for $y=T$),
we find that $\Gamma_{rel}^R=\Gamma_{rel}^T=\Gamma_{rel}$,
where
\begin{equation}
 \Gamma_{rel} = \frac{eV_{det}} {h} \Delta {\cal T} . \label{Relaxation-rate}
\end{equation}
\end{subequations}

Implications of these results (Eq.~\ref{eq:My}) are summarized as follows.
First, the charge state evolves into
 $|0\rangle$ ($|1\rangle$) with the relaxation rate $\Gamma_{rel}$
(Eq.(\ref{Relaxation-rate})) under
 continuous projection of detector electrons onto lead $T$($R$).
The direction of the evolution depends on which output lead
is selected. It is important to note that the conditional
state remains as a pure state as a result of measurement, in contrast to the case
of dephasing. We also point out that the conditional relaxation considered
here is different
from the stochastic evolution under random selection of measurement outcome
due to the partition noise of the QPC~\cite{korotkov99,averin06}.
In order to observe conditional relaxation under monitoring only
one particular output lead, we need to
correlate the state of the target system with that of the detector output.
(See below for observing this correlation.)
Second, $\Gamma_{rel}$ is much larger than $\Delta {\cal T}$-dependence $\Gamma_{dep}$. Because only one particular output is continuously selected,
 the conditional relaxation is not restricted by the shot noise
of the QPC detector, unlike the dephasing process. Finally, $\Gamma_{rel}$
depends only on $\Delta {\cal T}$, while $\Gamma_{dep}$ depends both on
$\Delta {\cal T}$ and $\Delta\phi$. Dephasing is related to the state
information transferred to the detector and therefore to the possibility of
measurement. On the other hand, by selecting one particular lead in the
detector, the phase part ($\Delta\phi$) of the state information is erased.
In fact, this behavior corresponds to the quantum erasure of the charge state
information encoded in the relative scattering phase $\Delta\phi$~\cite{kang07}.

\begin{figure}[b]
\includegraphics[width=75mm]{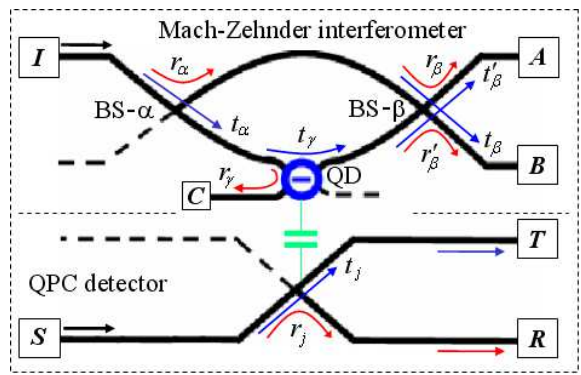}
 \caption{\label{fig2}
 A schematic diagram of a two-path interferometer coupled to a quantum point contact detector.
 Cross-correlation measurement between an output lead from the interferometer ($A$ or $B$) and the other
 from the detector ($T$ or $R$) reveals the nature of conditional relaxation. (See text for a discussion.) }
\end{figure}

Next, we propose a possible experiment to observe the effect of
the conditional relaxation. For a target system, we consider an
electronic two-path interferometer with a quantum dot (QD)
embedded in one of the two paths. The QD is capacitively coupled to a QPC
detector. (See Fig.~2.) The two-path interferometer can be
implemented by constructing a double-slit type Aharonov-Bohm
interferometer~\cite{schuster97}. Alternatively, it can be built up by
two beam splitters (BS-$\alpha$ and BS-$\beta$) with quantum Hall edge
state. This is an electronic analogue of the Mach-Zehnder
interferometer (MZI)~\cite{ji03}. The electronic transport in the
interferometer is characterized by the scattering matrix at
BS-$\alpha$, BS-$\beta$ and QD,
\begin{equation}
 S_i =  \left(
        \begin{array}{cc}
           r_i & t'_i \\
       t_i & r'_i
     \end{array}
       \right), \label{imz-smtx}
 \end{equation}
 where $i=\alpha,\beta,\gamma$. The reflection and the transmission probabilities are written as ${\cal R}_i = |r_i|^2$ and ${\cal T}_i = |t_i|^2$, respectively.

Because of the dwell time in the QD (denoted as $\Gamma^{-1}$), the
dephasing effect due to coupling to the QPC detector appears in
the probability ($P_x$) to find an electron at
lead $x$ ($x=A,B$)
\begin{subequations}
\begin{equation}
 P_x = \tr_{_{MZI}} \big[ c_x^{\dag}c_x\,\rho (t=\Gamma^{-1}) \big],
\label{eq:Px}
\end{equation}
 where $\tr_{MZI} [\cdots]$ sums over the MZI degree
of freedom. Eq.(\ref{eq:Px}) implies that the electron is (on
average) collected at lead $x$ after time $t=1/\Gamma$ upon
injection. it gives
\begin{eqnarray}
 P_A = {\cal R}_\alpha {\cal R}_\beta + {\cal T}_\alpha {\cal T}_\beta {\cal T}_\gamma + 2 {\cal{V}} M \cos(\varphi + \phi_\nu), \label{Prob-a}\\
 P_B = {\cal R}_\alpha {\cal T}_\beta + {\cal T}_\alpha {\cal R}_\beta {\cal T}_\gamma - 2 {\cal{V}} M \cos(\varphi + \phi_\nu), \label{Prob-b}
\end{eqnarray}
 where $\varphi = \arg( t_\alpha t_\gamma t'_\beta / r_\alpha r_\beta )$,
 $ M = ( {\cal R}_\alpha {\cal R}_\beta {\cal T}_\alpha {\cal T}_\beta {\cal T}_\gamma )^{1/2}$,
  and $\phi_\nu =eV_{det} \arg(\nu_{01}) /h \Gamma$.
  The visibility factor (${\cal V}$) depends on the dephasing rate of
Eq.(\ref{Dephasing-rate}) as
\begin{equation}
{\cal{V}} = e^{-\Gamma_{dep}/\Gamma} \simeq 1-\Gamma_{dep}/\Gamma,
\end{equation}
\end{subequations}
in the limit of $\Gamma_{dep}/\Gamma \ll 1$, which agrees with  previous
results~\cite{aleiner97,silva01}.

The conditional probability ($P_{x|y}$) to find an electron at
lead $x$ ($x = A,B$)
 conditioned on a particular detector output $y$ ($y= T,R$) is obtained from
the relation
 $P_{x|y} = \langle\Psi|M_y c_x^\dagger c_x M_y|\Psi\rangle = N_y^2 P'_{x|y}  $,
 where $N_y  =   \large[  \langle \Psi  |  y \rangle \langle  y| \Psi \rangle \large]^{-1/2}$
 is the normalization factor of the conditional state. In an experiment, $P'_{x|y}$ is the more relevant quantity for measurement. (See below.)
 At time $t=1/\Gamma$, it is given as
\begin{subequations}
\label{eq:P-prime}
\begin{eqnarray}
 P'_{A| y } = {\cal R}_\alpha {\cal R}_\beta  + e^{\pm\Gamma_{rel}/\Gamma} {\cal T}_\alpha {\cal T}_\beta {\cal T}_\gamma  + 2 {\cal{V}}_ y  M \cos\varphi_ y,\\
 P'_{B| y } =  {\cal R}_\alpha {\cal T}_\beta  + e^{\pm\Gamma_{rel}/\Gamma} {\cal T}_\alpha {\cal R}_\beta {\cal T}_\gamma - 2 {\cal{V}}_y   M \cos\varphi_ y,
\end{eqnarray}
 where $\varphi_{y} = \varphi+ \xi_y/\Gamma $ ($y=R,T$).
 The visibility factor that appears in the interference term is given by
\begin{equation}
   {\cal{V}}_ y =e^{\pm\Gamma_{rel}/2\Gamma} \cong 1 \pm  \frac {
  \Gamma_{rel}}{2\Gamma}.
\end{equation}
\end{subequations}
In Eq.(\ref{eq:P-prime}), $+$ ($-$) sign is for  $ y  = R$ ($T$).
In contrast to the case of dephasing in single-particle transport,
the visibility of the conditional probability can be enhanced (for $y=R$) or
reduced (for $y=T$) depending on which output lead is chosen in the detector.
  The enhancement (reduction) of the visibility for $y=R$ ($y=T$) is because selecting detector electrons at lead $R$ ($T$)
 effectively increases (decreases) the transmission probability through the QD.

In the following, we show that the cross-correlation measurement
of current at leads $x$ ($x= A,B$) and $y$ ($y= T,R$) is
directly related to the quantity $P'_{x|y}$ in
Eq.(\ref{eq:P-prime}). The bias voltage, $V$, applied to the MZI
is assumed to be much smaller than that of the detector: $V\ll
V_{det}$. The frequency-dependent current cross correlation
$S_{xy}(\omega)$ is defined by
\begin{eqnarray}
2 \pi \delta(\omega+\omega') S_{ x y }(\omega)
  ~~~~~~~~~~~~~~~~~~~~~~~~~~~~~~~~~~~~~\nonumber  \\
 =  \langle \bar{\Psi}|  \Delta {\cal{I}}_ x(\omega)\Delta{\cal{I}}_ y (\omega')
   + \Delta {\cal{I}}_ y (\omega')   \Delta{\cal{I}}_ x(\omega)|\bar{\Psi} \rangle,
\label{def-NPS}
\end{eqnarray}
 where $|\bar{\Psi}\rangle$ is the many-electron transport state of the composite system. $\Delta {\cal{I}}_l$ is the current fluctuation defined by
 $\Delta {\cal{I}}_l ={\cal{I}}_l - \langle {\cal{I}}_l \rangle$ where
 ${\cal I}_l$ is the output current operator at lead $l$.

In evaluating the expectation values in Eq.~(\ref{def-NPS}), we need to calculate quantities such as $\langle \bar{\Psi}| c_x^\dagger(E)c_x(E') c_y^\dagger(\epsilon)c_y(\epsilon') |\bar{\Psi}\rangle$. $E,E'$ and $\epsilon,\epsilon'$ are the energies of electrons injected from the interferometer and the detector, respectively. These energies are in the ranges
$0\leq E,E'\leq eV$ and $0\leq \epsilon,\epsilon'\leq eV_{det}$. In order to
calculate such quantities, we made the following assumptions: (i) All of the scattering matrices are independent of the energies. This assumption is valid as long as the bias voltages are not very large to alter the characteristics of the QPCs. (ii) The density matrix of the whole system, $\bar{\rho}\equiv |\bar{\Psi}\rangle \langle \bar{\Psi}|$ can be written
as a direct product
\begin{equation}
 \bar{\rho} \simeq \bar{\rho}_1\otimes\bar{\rho}_2 ,
\end{equation}
where $\bar{\rho}_1$ is the part of the density matrix that contains energies $E,E',\epsilon,\epsilon'$, while $\bar{\rho}_2$ represents the remaining part. This is a reasonable assumption because the different energy states
of electrons are unlikely to interfere with each other. Using these assumptions, we obtain
a simple relation of the zero-frequency cross-correlation
\begin{equation}
 S_{ x y }(0)= \frac{e^3}{\pi \hbar} V \big[P'_{ x| y }
 P_ {y0}  - P_ x P_ y  \big],
\end{equation}
where $P_{R0} =  {\cal R}_0 $, $ P_{T0} =  {\cal  T}_0 $,
  $P_{R} =  {\cal R}_\alpha {\cal R}_0 + {\cal T}_\alpha {\cal R}_1 $,
  and $P_{T} =  1- R_R$.
Also, it is straightforward to find that the average current
$\langle {\cal I}_x\rangle$ at lead $x$ satisfies the Landauer
formula: $\langle {\cal I}_x\rangle = (e^2/2\pi\hbar)P_xV$.
(Similarly, $\langle {\cal I}_y\rangle =
(e^2/2\pi\hbar)P_yV_{det}$ for the detector.)
 Therefore, analyzing the
cross correlation $S_{ x y }(0)$ as well as the DC currents reveals
the characteristic features of conditional relaxation
and dephasing.

 In conclusion, we have found that a linearly superposed charge state is conditionally relaxed
 under continuous measurement by an attached QPC detector. The direction of
 the relaxation depends on the choice of the detector output lead. It
 takes place much faster than the current-sensitive part of dephasing.
  We suggest that this feature can be revealed by constructing an interferometer for the charge state and investigating the current-current correlation between the two subsystems.

\acknowledgements%
This work was supported by the Korea Research Foundation
(KRF-2005-070-C00055) and by the ``Cooperative Research Program" of
the Korea Research Institute of Standards and Science.


\end{document}